\documentclass[11pt]{llncs}
\usepackage{fullpage}
\usepackage[dvips]{graphicx}
\begin{document}
%\baselineskip = 1.2\baselineskip
\title{On-line Viterbi Algorithm and Its Relationship to Random Walks}
\author{Rastislav \v{S}r\'amek\inst{1}
        \and Bro\v{n}a Brejov\'a\inst{2}
        \and Tom\'a\v{s} Vina\v{r}\inst{2}}
\institute{Department of Computer Science,
           Comenius University,\\842~48 Bratislava, Slovakia,
           e-mail: rasto@ksp.sk
           \and
           Department of Biological Statistics and Computational Biology,
           Cornell University,\\ Ithaca, NY 14853, USA,
           e-mail: \{bb248,tv35\}@cornell.edu}

\maketitle

\begin{abstract}
In this paper, we introduce the on-line Viterbi algorithm for decoding
hidden Markov models (HMMs) in much smaller than linear space.  Our
analysis on two-state HMMs suggests that the expected maximum memory
used to decode sequence of length $n$ with $m$-state HMM can be as low
as $\Theta(m\log n)$, without a significant slow-down compared to the
classical Viterbi algorithm. Classical Viterbi algorithm requires
$O(mn)$ space, which is impractical for analysis of long DNA sequences
(such as complete human genome chromosomes) and for continuous data
streams.  We also experimentally demonstrate the performance of
the on-line Viterbi algorithm on a simple HMM for gene finding on both
simulated and real DNA sequences.

\end{abstract}

\section{Introduction}

Hidden Markov models (HMMs) are generative probabilistic models that
have been succesfuly used for annotation of sequence data, such as DNA and
protein sequences, natural langauge texts, and sequences of observations or
measurements. Their numerous applications include gene finding
\cite{Burge1997}, protein secondary structure prediction
\cite{Krogh2001}, and speech recognition \cite{Rabiner1989}.
The linear-time Viterbi algorithm \cite{Forney1973} is the most commonly
used algorithm for these tasks.  Unfortunately, the space required by
the Viterbi algorithm grows linearly with the length of the sequence
(with a high constant factor), which makes it unsuitable 
for analysis of continuous or very long sequences. For example, 
DNA sequence of a single chromosome can be hundreds of megabases long.
In this paper, we address this
problem by proposing an on-line Viterbi algorithm that on average
requires much less memory and that can annotate
continuous streams of data on-line without reading the complete
input sequence first. 

An HMM, composed of states and transitions, is a probabilistic model
that generates sequences over a given alphabet. In each step of this
generative process, the current state generates one symbol of the
sequence according to the \emph{emission probabilities} associated
with that state.  Then, an outgoing transition is randomly chosen
according to the \emph{transition probability table}, and this
transition is followed to the new state. This process is repeated
until the whole sequence is generated.

The states in the HMM represent distinct features of the observed
sequences (such as protein coding and non-coding sequences in a
genome), and the emission probabilities in each state represent
statistical properties of these features. The HMM thus defines a joint
probability $\Pr(X,S)$ over all possible sequences $X$ and all
\emph{state paths} $S$ through the HMM that could generate these
sequences. To annotate a given sequence $X$, we want to recover the
state path $S$ that maximizes this joint probability. For example, in 
an HMM with one state for protein-coding sequences, and one
state for non-coding sequences, the most probable state path marks
each symbol of the input sequence $X$ as either protein coding or
non-coding.

To compute the most probable state path, we use the Viterbi dynamic
programming algorithm \cite{Forney1973}. For every prefix $X_1\dots X_i$
of the given sequence $X$ and for every state $j$, we compute 
the most probable state path generating this prefix ending in state 
$j$. We store the probability of this path in table $P(i,j)$
and its second last state in table $B(i,j)$. These values 
can be computed from left to right, using the recurrence
$P(i,j)=\max_k\{P(i-1,k)\cdot t_k(j) \cdot e_j(X_i)\}$, where $t_k(j)$
is the transition probability from state $k$ to state $j$, and
$e_j(X_i)$ is the emission probability of the $i$-th symbol of X in
state $j$. Back pointer $B(i,j)$ is the value of $k$ that maximizes
$P(i,j)$. After computing these values, we can recover the most
probable state path $S=s_1,\allowbreak{}\dots,\allowbreak{}s_n$ by
setting the last state as $s_n=\arg\max_k\{P(n,k)\}$, and then
following the back pointers $B(i,j)$ from right to left (i.e., $s_i =
B(i+1,s_{i+1})$). For an HMM with $m$ states and a sequence $X$ of length $n$, 
the running time of the Viterbi algorithm is $\Theta(nm^2)$, 
and the space is $\Theta(nm)$.

This algorithm is well suited for sequences and models of moderate size.
However, to annotate all 250 million symbols of the human chromosome 1
with a gene finding HMM consisting of hundred states, 
we would require 25~GB of memory just to store the
back pointers $B(i,j)$. This is clearly impractical on
most computational platforms.

Several solutions are used in practice to overcome this problem. 
For example, most
practical gene finding programs process only sequences of limited
size. The long input sequence is split into several shorter sequences
which are processed separately. Afterwards, the results are merged
and conflicts are resolved heuristically. This approach leads to 
suboptimal solutions, especially if the genes we are looking for
cross the boundaries of the split.

Grice et al. \cite{Grice1997} proposed 
a practical checkpointing algorithm that trades running time
for space. We divide the
input sequence into $K$ blocks of $L$ symbols, and during the forward
pass, we only keep the first column of each block. To obtain the most
probable state path, we recompute the last block of $L$ columns, and use back
pointers to recover the last $L$ states of the most probable path, as well
as the last state of the previous block. The information about this
last state can now be used to recompute the most probable state path
within the previous block in the same way, and the process is repeated
for all blocks. Since every value of $P(i,j)$
will be computed twice, this means two-fold slow-down compared to 
the Viterbi algorithm, but if we set $K=L=\sqrt{n}$, this algorithm
only requires $\Theta(\sqrt{n}m)$ memory. Checkpointing can
be further generalized to trade $L$-fold slow-down for memory
of $\Theta(\sqrt[L]{n}m)$ \cite{Tarnas1998,Wheeler2000}.

In this paper, we propose and analyze an on-line Viterbi algorithm that
does not use fixed amount of memory for a given sequence.
Instead, the amount of memory varies depending on the properties of the
HMM and the input sequence. In the worst case, our algorithm still
requires $\Theta(nm)$ memory; however, in practice the requirements
are much lower.  We prove, by demonstrating analogy
to random walks and using results from the theory of extreme values, that
in simple cases
the expected space for a sequence of length $n$ is as low as
$\Theta(m\log n)$. We also experimentally demonstrate that the memory
requirements are low for more complex HMMs.

%%%%%%%%%%%%%%%%%%%%%%%%%%%

\section{On-line Viterbi algorithm}

\begin{figure}[t]
\centerline{\includegraphics[scale=0.8]{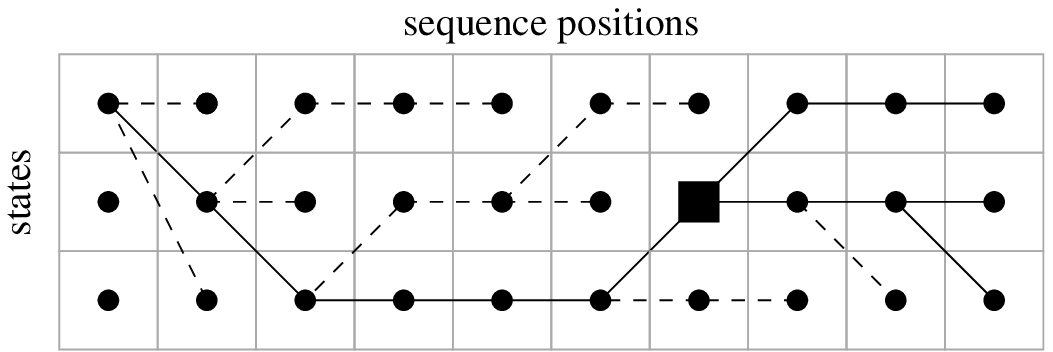}}
\caption{{\bf Example of the back pointer tree structure.} 
Dashed lines mark the edges that
cannot be part of the most probable state path. The square node marks
the coalescence point of the remaining paths.}
\label{fig:trellis}
\end{figure}

In our algorithm, we represent the back
pointer matrix $B$ in the Viterbi algorithm by a tree structure 
(see \cite{Forney1973}), with node $(i,j)$ for each sequence position
$i$ and each state $j$. Parent of node $(i,j)$ is 
the node $(i-1,B(i,j))$. In this 
data structure, the most probable state path is a path from the
leaf node $(n,j)$ with the highest probability $P(n,j)$ 
to the root of the tree (see Figure \ref{fig:trellis}).

This tree is built as the Viterbi algorithm progresses from
left to right. After processing sequence position $i$, all 
edges that do not lie on one of the paths ending in a level
$i$ node can be removed; these edges will not be used in the most
probable path \cite{Henderson1997}. The remaining $m$ paths represent
all possible initial segments of the most probable state path. These
paths are not necessarily edge disjoint; in fact, often all the
paths share the same prefix up to some node called \emph{coalescence point} 
(see Figure \ref{fig:trellis}).

Left of the coalescence point, there is only a single candidate for
the initial segment of the most probable state path. Therefore we can
output this segment and remove all edges and nodes of the tree up
to the coalescence point. Forney \cite{Forney1973} describes an
algorithm that after processing $D$ symbols of the input sequence checks
whether a coalescence point has been reached; in such case, the
initial segment of the most probable state path is outputted.  If the
coalescence point was not reached, one potential initial segment is
chosen heuristicaly. Several studies \cite{Hemmati1977,Onyszchuk1991} 
suggest how to choose
$D$ to limit the expected error caused by such heuristic steps 
in the context of convolution codes.
 
Here we show how to detect the existence of a coalescence point
dynamically without introducing significant overhead to the whole
computation. We maintain a compressed version of the back pointer tree,
where we omit all internal nodes that have
less than two children. Any path consisting of such nodes will be
contracted to a single edge.  This compressed tree has $m$ leaves
and at most $m-1$ internal nodes. Each node stores the number of its
children and a pointer to its parent node. 
We also keep a linked list of all the
nodes of the compressed tree ordered by the sequence
position. Finally, we also keep the list of pointers to all the
leaves.

When processing the $k$-th sequence position in the Viterbi algorithm, we
update the compressed tree as follows. First, we create a new leaf
for each node at position $i$, link it to its parent (one of the
former leaves), and insert it into the linked list. Once these new
leaves are created, we remove all the former leaves that have no
children, and recursively all of their ancestors that would not have
any children. 

Finally, we need to compress the new tree: we examine all the nodes
in the linked list in order of decreasing sequence position.  If the
node has zero or one child and is not a current leaf, we simply delete
it. For each leaf or node that has at least two children, we follow
the parent links until we find its first ancestor (if any) that has at
least two children and link the current node directly to that
ancestor. A node $(\ell,j)$ that does not have an ancestor with at least
two children is the coalescence point; it will become a new root.
We can output the most probable state path for all sequence positions
up to $\ell$, and remove all results of computation for these
positions from memory.

The running time of this update is $O(m)$ per sequence position, and
the representation of the compressed tree takes $O(m)$ space. Thus
the asymptotic running time of the Viterbi algorithm is not increased 
by the maintanance of the compressed tree. Moreover, we have implemented both
the standard Viterbi algorithm and our new on-line extension, and
the time measurements suggest that the overhead required for the
compressed tree updates is less than 5\%.

The worst-case space required by this algorithm is still
$O(nm)$. However, this is rarely the case for realistic data; required
space changes dynamically depending on the input.
In the next section, we show that for simple HMMs the expected maximum
space required for processing sequence of length $n$ is
$\Theta(m\log n)$. This is much better than checkpointing, which
requires space of $\Theta(m\sqrt{n})$ with a significant increase
in running time. We conjecture that this trend extends to more
complex cases. We also present experimental results on a 
gene finding HMM and real DNA sequences showing that the on-line
Viterbi algorithm leads to significant savings in memory.

Another advantage of our algorithm is that it can construct initial
segments of the most probable state path before the whole input
sequence is read. This feature makes it ideal for on-line processing 
of signal streams (such as sensor readings).

\section{Memory requirements of the on-line Viterbi algorithm}

In this section, we analyze the memory requirements of the on-line Viterbi
algorithm. The memory used by the algorithm is variable throughout the
execution of the algorithm, but of special interest are asymptotic
bounds on the expected maximum amount of memory used by the algorithm
while decoding a sequence of length $n$.

We use analogy to random walks and results in extreme value theory to
argue that for a symmetric two-state HMMs, the expected maximum memory
is $\Theta(m\log n)$. We also conduct experiments on an HMM for 
gene finding, and both real and simulated DNA sequences.

\subsection{Symmetric two-state HMMs}

\begin{figure}[t]
\centerline{\includegraphics[width=0.8\textwidth]{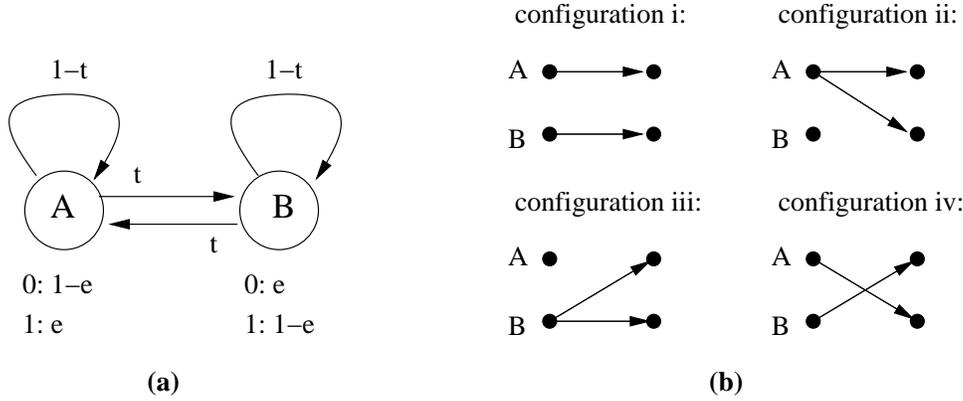}}
\caption{{\bf (a) Symmetric two-state HMM} with two parameters: 
$e$ for emission
probabilities and $t$ for transitions probabilities. 
{\bf (b) Possible back-pointer configurations} for the two-state HMM.
\label{fig:twostate}}
\end{figure}

Consider a two-state HMM over a binary alphabet as shown in Figure
\ref{fig:twostate}a. For simplicity, we assume $t<1/2$ and $e<1/2$.
The back pointers between the sequence positions $i$ and $i+1$ can
form one of the configurations i--iii shown in Figure
\ref{fig:twostate}b. Denote $p_A=\log P(i,A)$ and $p_B=\log P(i,B)$,
where $P(i,j)$ is the table of probabilities from the Viterbi algorithm.
The recurrence used in the Viterbi algorithm implies that 
the configuration i occurs when $\log t-\log(1-t)\le p_A-p_B\le \log (1-t)
- \log t$, configuration ii occurs when $p_A-p_B\ge \log(1-t)-\log t$, 
and configuration iii occurs when $p_A-p_B\le \log t - \log(1-t)$.
Configuration iv never happens for $t<1/2$.

Note that for a two-state HMM, a coalescence point occurs
whenever one of the configurations ii or iii occur. Thus the memory
used by the HMM is proportional to the length of continuous sequence
of configurations i. We will call such a sequence of configurations
a \emph{run}.

First, we  analyze the length distribution of runs under the
assumption that the input sequence $X$ is a sequence of uniform
i.i.d. binary random variables. In such case, we represent the run 
by a symmetric random walk corresponding
to a random variable 
$X=\frac{p_A-p_B}{\log (1-e) - \log e} - (\log t-\log(1-t)).$ Whenever
this variable is within the interval $(0,K)$, where 
$K = \left\lceil 2 \frac{\log(1-t)-\log(t)}{\log(1-e)-\log(e)}\right\rceil,$
the configuration i occurs, and the quantity $p_A-p_B$ is updated by 
$\log(1-e)-\log e$, if the symbol at the corresponding sequence position
is 0, or $\log e - \log(1-e)$, if this symbol is 1. These shifts
correspond to updating the value of $X$ by $+1$ or $-1$. 

When $X$ reaches 0, we have a coalescence point in configuration iii, and
the $p_A-p_B$ is initialized to $\log t - \log(1-t) \pm (\log e - \log 1-e)$,
which either means initialization of $X$ to $+1$, or another coalescence
point, depending on the symbol at the corresponding sequence position.
The other case, when $X$ reaches $K$ and we have a coalescence point in
configuration ii, is symmetric. 

We can now apply the classical results from the theory of random walks
(see \cite[ch.14.3,14.5]{Feller1968}) to analyze the expected length
of runs.

\begin{lemma}
Assuming that the input sequence is uniformly i.i.d., the expected length of a
run of a symmetrical two-state HMM is $K-1$.
\end{lemma}

Therefore the larger is $K$, the more memory is required to decode the
HMM. The worst case is achieved as $e$ approaches $1/2$.
In such case, the two states are indistinguishable and being in state
$A$ is equivalent to being in state $B$. Using the theory of random walks,
we can also characterize the distribution of length of runs.

\begin{lemma}
\label{lem:distrib}
Let $R_\ell$ be the event that the length of a run of a symmetrical
two-state HMM is either $2\ell+1$ or $2\ell+2$. Then,
assuming that the input sequence is uniformly i.i.d., for some constants
$b,c>0$:
\begin{equation}
b\cdot\cos^{2\ell}\frac{\pi}{K}\le \Pr(R_\ell) 
\le c\cdot \cos^{2\ell}\frac{\pi}{K}
\end{equation}
\end{lemma}

\def\pivk{\frac{\pi v}{K}}
\def\pik{\frac{\pi}{K}}
\begin{proof}
For a symmetric random walk on interval $(0,K)$ with absorbing barriers 
and with starting point
$z$, the probability of event $W_{z,n}$ that this random walk ends
in point $0$ after $n$ steps is zero, if $n-z$ is odd, and the
following quantity, if $n-z$ is even \cite[ch.14.5]{Feller1968}:
\begin{equation}
\Pr(W_{z,n}) = \frac{2}{K}\sum_{0<v<K/2} 
       \cos^{n-1}\pivk \sin\pivk \sin\frac{\pi z v}{K}
\end{equation}
Using symmetry, note that the probability of the same random walk
ending after $n$ steps at barrier $K$ is the same as probability of
$W_{K-z,n}$. Thus, if $K$ is odd, we can state:
\begin{eqnarray}
\Pr(R_\ell) &=& \Pr(W_{1,2\ell+1}) + \Pr(W_{K-1,2\ell+1}) \nonumber\\
            &=& \frac{2}{K}\sum_{0<v<K/2}\cos^{2\ell}\pivk
              \sin\pivk\left(\sin\pivk+(-1)^{v+1}\sin\pivk\right)
              \nonumber\\
            &=& \frac{4}{K}\sum_{0<v<K/2,\mbox{ $v$ odd}}
                  \cos^{2\ell}\pivk\sin^2\pivk
\end{eqnarray}
There are at most $K/4$ terms in the sum and they can all be bounded from above
by 
$\cos^{2\ell}\pivk$. Thus, we can
give both upper and lower bounds on $\Pr(R_\ell)$ using only the 
first term of the sum as follows:
\begin{equation}
\frac{4}{K}\sin^2\pik \cos^{2\ell}\pik
\le \Pr(R_\ell) \le \cos^{2\ell}\pik
\end{equation}
Similarly, if $K$ is even, we can state:
\begin{eqnarray}
\Pr(R_\ell) &=& \Pr(W_{1,2\ell+1}) + \Pr(W_{K-1,2\ell+2})\nonumber \\
            &=& \frac{2}{K}\sum_{0<v<K/2}\cos^{2\ell}\pivk
                           \sin^2\pivk\left(1+(-1)^{v+1}\cos\pivk\right)
\end{eqnarray}
and thus we have a similar bound:
\begin{equation}
\frac{2}{K}\sin^2\pik\left(1+\cos\pik\right)\cos^{2\ell}\pik
\le \Pr(R_\ell) \le 2\cos^{2\ell}\pik
\end{equation}
\qed
\end{proof}

The previous lemma characterizes the length distribution of a single
run. However, to analyze memory requirements for a sequence of length
$n$, we need to consider maximum over several runs whose total length
is $n$.  Similar problem was studied for the runs of
heads in a sequence of $n$ coin tosses
\cite{Guibas1980,Gordon1986}. For coin tosses, the length distribution
of runs is geometric, while in our case the runs are only bounded by
geometricaly decaying functions. Still, we can prove that the expected
length of the longest run grows logarithmically with the length of the
sequence, as is the case for the coin tosses.

\begin{lemma}
\label{lem:max}
Let $X_1,X_2,\dots$ be a sequence of i.i.d. random variables drawn from a
geometrically decaying distribution over positive integers, i.e. 
there exist constants $a,b,c$, $a\in (0,1)$,
$0<b\le c$, such that for all integers $k\ge 1$, 
$b a^k \le \Pr(X_i > k) \le c a^k.$

Let $N_n$ be the largest index such that $\sum_{i=1\dots N_n} X_i\le n$,
and let $Y_n$ be $\max\{X_1,X_2,\dots,X_{N_n},n-\sum_{i=1}^{N_n} X_i\}$.
Then
\begin{equation}
E[Y_n] = \log_{1/a} n + o(\log n)
\end{equation}
\end{lemma}

\begin{proof}

Let $Z_n = \max_{i=1\dots n} X_n$ be the maximum of the first $n$
runs. Clearly, $\Pr(Z_n \le k) = \Pr(X_i \le k)^n$, and therefore
$(1 - c a^k)^n \le \Pr(Z_n \le k) \le (1 - b a^k)^n$ for all integers
$k\ge \log_{1/a}(c)$.

\paragraph{Lower bound:}
Let $t_n = \log_{1/a} n - \sqrt{\ln n}$. 
If $Y_n\le t_n$, we need at
least $n/t_n$ runs to reach the sum $n$, i.e.
$N_n\ge n/t_n-1$ (discounting the last
incomplete run). Therefore
\begin{equation}
\Pr(Y_n\le t_n) \le \Pr(Z_{\frac{n}{t_n}-1} \le t_n) 
\le (1 - b a^{t_n})^{\frac{n}{t_n}-1}=
(1-ba^{t_n})^{a^{-t_n}a^{t_n}(\frac{n}{t_n}-1)}
\end{equation}

Since $\lim_{n \to \infty} a^{t_n}(n/t_n-1) =
\infty$ and $\lim_{x \to 0} (1-b x)^{1/x} =
e^{-b}$, we get $\lim_{n\to\infty} \Pr(Y_n\le t_n) = 0$.
Note that $E[Y_n] \ge t_n (1-\Pr(Y_n \le t_n))$, and thus we 
get the desired bound.

\paragraph{Upper bound:} 
Clearly, $Y_n\le Z_n$ and so $E[Y_n] \le E[Z_n]$.  
Let $Z'_n$ be the
maximum of $n$ i.i.d. geometric random variables $X'_1, \dots, X'_n$
such that $\Pr(X'_i\le k) = 1-a^k$.

We will compare
$E[Z_n]$ to the expected value of variable $Z'_n$.
Without loss of generality, $c\ge 1$.  For any real
$x\ge \log_{1/a}(c)+1$ we have:
\begin{eqnarray*}
\Pr(Z_n\le x) 
&\ge& (1-c a^{\lfloor x\rfloor})^n \\
&=& \left(1-a^{\lfloor x\rfloor -\log_{1/a}(c)}\right)^n\\
&\ge& \left(1-a^{\lfloor x -\log_{1/a}(c)-1\rfloor}\right)^n\\
&=& \Pr(Z'_n\le x -\log_{1/a}(c)-1)\\
&=& \Pr(Z'_n+\log_{1/a}(c)+1 \le x)
\end{eqnarray*}
This inequality holds even for $x<\log_{1/a}(c)+1$, since the
right-hand side is zero in such case.
Therefore, $E[Z_n]\le E[Z'_n+\log_{1/a}(c)+1] =E[Z'_n] + O(1)$.
Expected value of $Z'_n$ is $\log_{1/a}(n)+o(\log n)$ \cite{Schuster1985},
which proves our claim.\qed
\end{proof}

%% sum_i=k^infty a^k = a^k/(a-1) (to apply distributions)
%% need to multiply by two, since this is a distribution
%% of 2-steps rather than single steps
%% 2*1/ln(1/cos^2(\pi/K)) = 1/ln(1/cos(\pi/K))

Using results of Lemma \ref{lem:max} together with the
characterization of run length distributions by Lemma
\ref{lem:distrib}, we can conclude that for symmetric two-state HMMs,
the expected maximum memory required to process 
a uniform i.i.d. input sequence of length $n$ is
$(1/\ln(1/\cos(\pi/K)))\cdot \ln n + o(\log n)$. \footnote{%
We omitted the first run, which has a different
starting point and thus does not follow the distribution
outlined in Lemma \ref{lem:distrib}. However, the expected
length of this run does not depend on $n$ and thus contributes only
a lower-order term. We also omitted the runs of length one that start
outside the interval $(0,K)$; these
runs again contribute only to lower order terms of the lower bound.}
Using the Taylor
expansion of the constant term as $K$ grows to infinity, 
$1/\ln(1/\cos(\pi/K))) = 2K^2/\pi^2 + O(1)$, 
we obtain that the maximum memory grows
approximately as $(2K^2/\pi^2)\ln n$.

The asymptotic bound $\Theta(\log n)$ can be easily extended to the
sequences that are generated by the symmetric HMM, instead of uniform
i.i.d. The underlying process can be described as a random walk with
approximately $2K$ states on two $(0,K)$ lines, each line
corresponding to sequence symbols generated by one of the two
states. The distribution of run lengths still decays geometrically
as required by Lemma \ref{lem:max}; the base of the exponent is the
largest eigenvalue of the transition matrix
with absorbing states omitted (see e.g. \cite[Claim 2]{Buhler2005}).

The situation is more complicated in the case of non-symmetric 
two-state HMMs.
Here, our random walks proceed in steps that are arbitrary real
numbers, different in each direction. We are not aware of any
results that would help us to directly analyze distributions
of runs in these models, however we conjecture that the size of
the longest run is still $\Theta(\log n)$. Perhaps, to obtain
bounds on the length distribution of runs, one can approximate
the behaviour of such non-discrete random walks by a different
model (for example, \cite[ch.7]{Durrett1996}).

\subsection{Multi-state HMMs}

\begin{figure}[t]
\begin{minipage}[b]{0.48\textwidth}
\centerline{\bf (a)}
\includegraphics[scale=0.55]{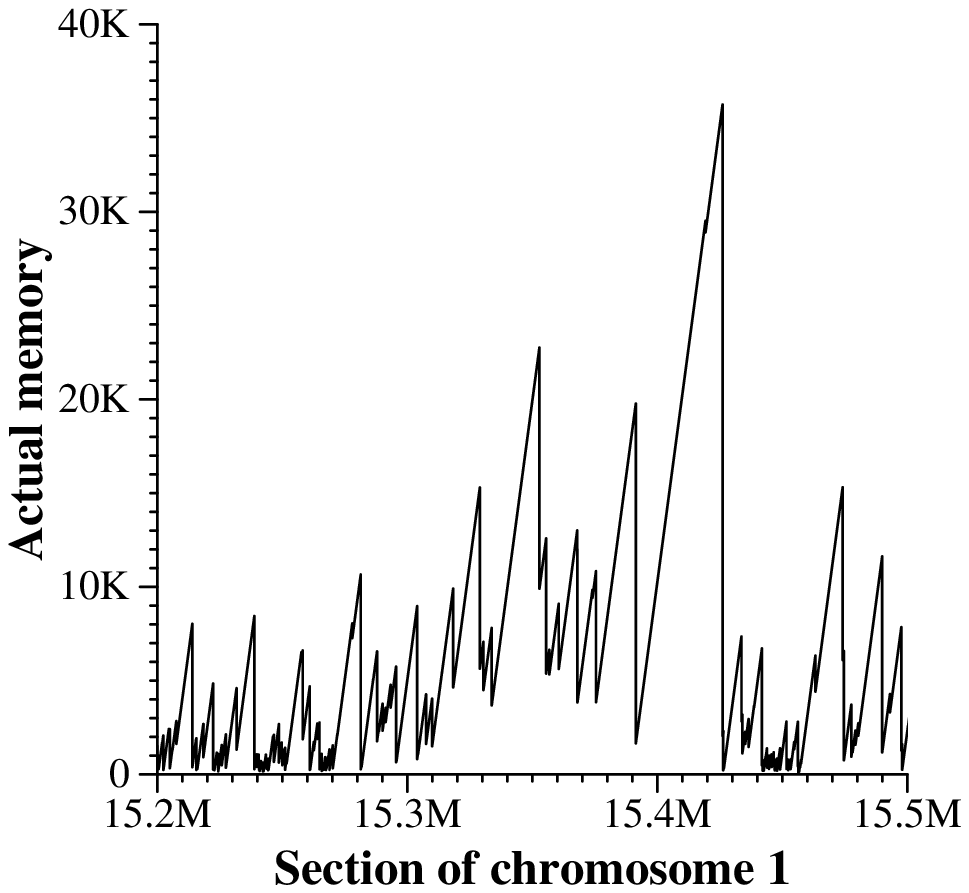}
\end{minipage}
\hfill
\begin{minipage}[b]{0.48\textwidth}
\centerline{\bf (b)}
\includegraphics[scale=0.55]{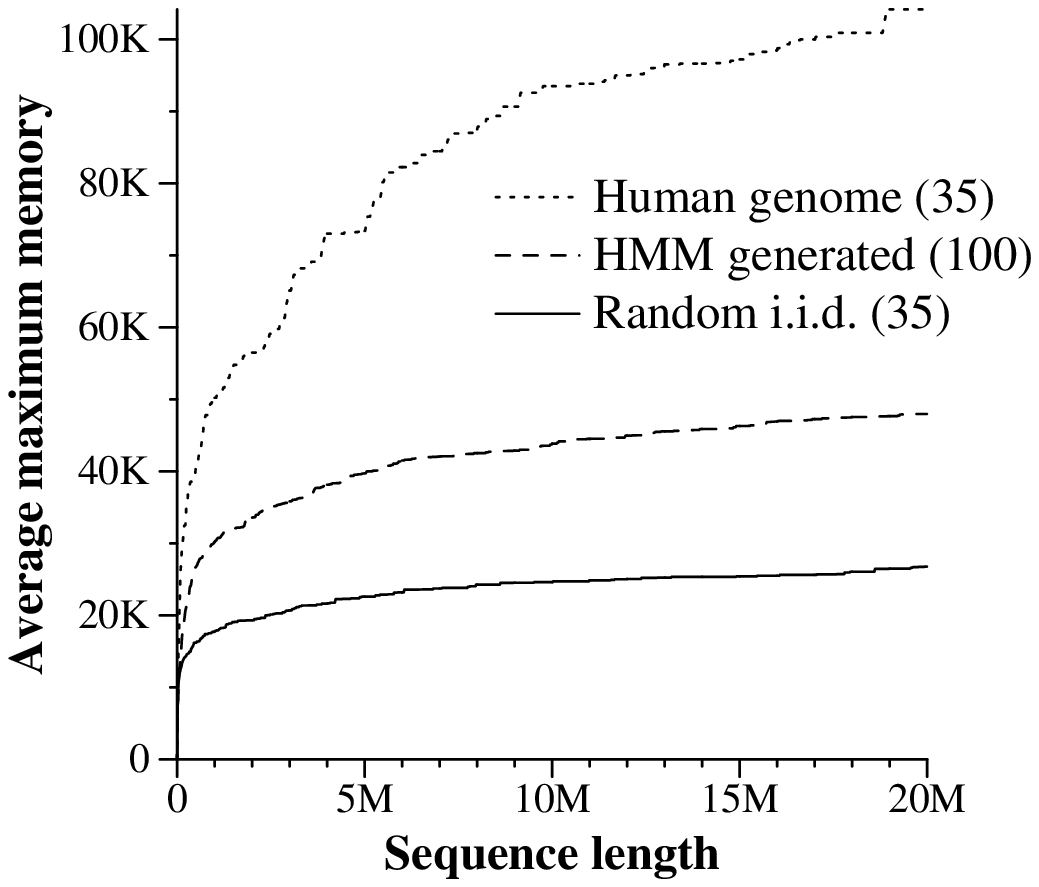}
\end{minipage}
\caption{{\bf Memory requirements of a gene finding HMM.} a) Actual
length of table used on a segment of human chromosome 1. b) Average maximum
table length needed for prefixes of 20~MB sequences.
\label{fig:max}} 
\end{figure}

Our analysis technique cannot be easily extended to HMMs with many
states. In two-state HMMs, each new coalescence event clears the
memory, and thus the execution of the algorithm can be divided
into more or less independent runs. A coalescent event in 
a multi-state HMM results in a non-trivial tree left in memory,
sometimes with a substantial depth. Thus, the sizes of 
consecutive runs are no longer independent
(see Figure \ref{fig:max}a).

To evaluate the memory requirements of our algorithm for multi-state
HMMs, we have implemented the algorithm and performed several experiments
on both simulated and biological sequences. First, we generalized
the symmetric HMMs from the previous section to multiple states.
The symmetric HMM with $m$ states emits symbols over $m$-letter
alphabet, where each state emits one symbol with higher probability
than the other symbols. The transition probabilities are equiprobable,
except for self-transitions. We have tested the algorithm for 
$m\le 6$ and sequences generated both by a uniform i.i.d. process, and
by the HMM itself. Observed data are consistent with the logarithmic
growth of average maximum memory needed to decode a sequence of length $n$
(data not shown).

We have also evaluated the algorithm using a simplified
HMM for gene finding with 265 states. The emission probabilities 
of the states
are defined using at most 4-th order Markov chains, and the
structure of the HMM reflects known properties of genes (similar
to the structure shown in \cite{Brejova2007}). The HMM was
trained on RefSeq annotations of human chromosomes 1 and 22.

In gene finding, we segment the input DNA sequence into exons
(protein-coding sequence intervals), introns (non-coding sequence
separating exons within a gene), and intergenic regions (sequence
separating genes). Common measure of accuracy is exon sensitivity (how
many of real exons we have succesfuly and exactly predicted). 
The implementation
used here has exon sensitivity 37\% on testing set
of genes by Guigo et al. \cite{Guigo2006}. A realistic gene finder,
such as ExonHunter \cite{Brejova2005}, trained on the same data set
achieves sensitivity of 53\%. This difference is due to additional features
that are not implemented in our test, namely GC content levels, non-geometric
length distributions, and sophisticated signal models.

\iffalse
masked sequence results         this           Genscan        ExonHunter
Gene Sensitivity                6.76%           15.88%          8.78%
Gene Specificity                3.13%           9.81%           12.50%
Exon Sensitivity                37.13%          58.84%          52.91%
Exon Specificity                29.27%          45.43%          66.47%
Nucleotide Sensitivity          71.48%          83.68%          77.55%
Nucleotide Specificity          36.62%          59.70%          80.13%
\fi

We have tested the algorithm on 20~MB long sequences: regions from the human
genome, simulated
sequences generated by the HMM, and i.i.d. sequences.
Regions of the human genome were chosen from hg18 assembly so that
they do not contain sequencing gaps. The distribution for the i.i.d. sequences
mirrors the distribution of bases in the human chromosome 1.

The results are shown in Figure \ref{fig:max}b. 
The average maximum length of the table over several samples appears to grow
faster than logarithmically with the length of the sequence, though
it seems to be bounded by a polylogarithmic function. It is not clear whether
the faster growth is an artifact that would disapear 
with longer sequences or higher number of samples. 

The HMM for gene finding has a special structure,
with three copies of the state for introns that have the same emission
probabilities and the same self-transition probability. 
In two-state symmetric HMMs, similar emission probabilities of the two states
lead to increase in the length of individual runs. Intron states of a
gene finder are an extreme example of this phenomenon.

Nonetheless, on average a table of length roughly 100,000 is sufficient to
to process sequences of length 20~MB, which is a 200-fold improvement compared
to the trivial Viterbi algorithm. In addition, the length of 
the table did not exceed
222,000 on any of the 20MB human segments. 
As we can see in Figure \ref{fig:max}a, most of the time the 
program keeps only relatively short table; the average length on the human
segments is 11,000. The low average length can be
of a significant advantage if multiple processes share the same memory.

\section{Conclusion} 

In this paper, we introduced the on-line Viterbi algorithm. Our
algorithm is based on efficient detection of coalescence points in
trees representing the state-paths under consideration of the dynamic
programming algorithm. The algorithm requires variable space that
depends on the HMM and on the local properties of the analyzed
sequence.  For two-state symmetric HMMs, we have shown that the
expected maximum memory used for analysis of sequence of length $n$ is
approximately only $(2K^2/\pi ^2)\ln n$. Our experiments on both
simulated and real data suggest that the asymptotic bound $\Theta(m\ln
n)$ also extend to multi-state HMMs, and in fact, for most of the time
throughout the execution of the algorithm, much less memory is used.

Further advantage of our algorithm is that it can be used for on-line
processing of streamed sequences; all previous algorithms that are
guaranteed to produce the optimal state path require the whole
sequence to be read before the output can be started.

There are still many open problems. We have only been
able to analyze the algorithm for two-state HMMs, though trends
predicted by our analysis seem to generalize even to more complex
cases. Can our analysis be extended to multi-state HMMs? Apparently,
design of the HMM affects the memory needed for the decoding algorithm;
for example, presence of states with similar emission probabilities
tends to increase memory requirements. Is it possible to characterize
HMMs that require large amounts of memory to decode? Can we characterize
the states that are likely to serve as coalescence points?

\paragraph{Acknowledgments:} Authors would like
to thank Richard Durrett for useful discussions. Recently, we have
found out that parallel work on this problem is also performed by
another research group \cite{Keibler2006}. Focus of their work is on
implementation of an algorithm similar to our on-line Viterbi
algorithm in their gene finder, and possible applications to
parallelization, while we focus on the expected space analysis.

\bibliographystyle{splncs} \bibliography{hmmmem}
\end{document}